\documentclass[12pt]{article}
\usepackage{pdproc,axodraw} 

\def\beq{\begin{equation}}
\def\eeq{\end{equation}}
\def\bea{\begin{eqnarray}}
\def\eea{\end{eqnarray}}

 \def \lsim{\mathrel{\vcenter
     {\hbox{$<$}\nointerlineskip\hbox{$\sim$}}}}
\def \gsim{\mathrel{\vcenter
     {\hbox{$>$}\nointerlineskip\hbox{$\sim$}}}}
\def\gappeq{\mathrel{\rlap {\raise.5ex\hbox{$>$}}
{\lower.5ex\hbox{$\sim$}}}}
\def\lappeq{\mathrel{\rlap{\raise.5ex\hbox{$<$}}}}

\def\CPv{{\begin{picture}(13,0)(0,0)\put(0,0){\rm
CP}\put(0,0){\line(2,1){15}}\end{picture}}~}

\begin{document}

\renewcommand{\thefootnote}{\alph{footnote}}
  
\title{ Flavoured Leptogenesis}

\author{ Sacha Davidson }

\address{ IPN de Lyon, Universit\'e  Lyon 1, CNRS,  Villeurbanne,  69622 cedex France}

\abstract{Thermal leptogenesis, in the seesaw model,
 is a popular mechanism for generating the
Baryon Asymmetry of the Universe.
It was noticed recently, that  including  lepton flavour 
can modify significantly the results.   These proceedings
aim to discuss  why and when
flavour matters,  in the thermal leptogenesis
scenario for hierarchical right-handed neutrinos. 
No Boltzmann Equations are introduced.}
   
\normalsize\baselineskip=15pt

\section{The Baryon Asymmetry of the Universe}

The Standard Model of particle physics 
is extraordinarily successful. However, among the
few observations it cannot explain, is the Baryon
Asymmetry of the Universe\cite{Dolgov:1991fr}: it is observed
that locally, in our patch  of the Universe,
there is an excess of visible matter (baryons) over anti-matter,
and this excess must extend  to the whole observable
Universe, because  we do not see gamma rays
from proton anti-proton annihilation \cite{Cohen:1997ac}.
This visible matter excess implies a baryon 
asymmetry. The magnitude of  
 a lepton asymmetry is unclear, 
 because there could be an
asymmetry stored in the ubiquitous cosmic
background of neutrinos.

\subsection{Observations}

The  amount of baryonic matter in the Universe affects
the fluctuations in the Cosmic Microwave Background.
WMAP data\cite{Spergel:2006hy}
 gives the excess of baryons $B$ over anti-baryons
$\bar{B}$, normalised to the density of photons today
:
\beq
\frac{n_B- n_{\bar{B}}}{n_{\gamma }} {\Big |}_{today}  = 
\frac{n_B- n_{\bar{B}}}{s} \frac{s_0}{n_{\gamma 0}}
 =  6.15 \pm 0.25  \times 10^{-10}
\label{YB}
\eeq 
It is convenient
to calculate $Y_B$,  the   baryon asymmetry
 relative to the entropy density $s$,
because $s =g_* \frac{2 \pi^2}{45} T^3$ 
is conserved during Universe expansion
\footnote{$g_*$ is the number of relativistic
degrees of freedom in thermal equilibrium.
 Today,  $g_* = 2 + 2 \frac{4}{11} 3$ accounts
for the entropy in photons and neutrinos, so that
the entropy today $s_0$  is 
$ \sim 7  n_{\gamma 0}$, where
$n_\gamma = 2 \frac{\zeta (3)}{\pi^2}T^3$  and  $\zeta(3) \simeq 1.2$.
}.  
The production of light nuclei by
Big Bang Nucleosynthesis also depends on $Y_B$\cite{Steigman:2005uz},
and the observed light element abundances are
consistent with  eqn (\ref{YB}). 

Before embarking on a study of how to
generate a baryon asymmetry, one could first wonder if
the Universe was  born with  this excess. 
However, this is incompatible with inflation,
during which a primordial asymmetry would have been
diluted to  irrelevance. And since inflation is
currently the only candidate to explain the
temperature fluctuations in the CMB, coherent
across many horizons, one can conclude that 
the baryon asymmetry must be generated after inflation.

\subsection{Ingredients}

The three  ingredients required to produce 
a Baryon Asymmetry were given by Sakharov \cite{Sakharov:1967dj}:\\
$\bullet$ baryon number violation ---required
to evolve from a state with $B = 0 $ to $B \neq 0$.\\
$\bullet$C and CP violation---particles and anti-particles
should behave differently,  to obtain  an asymmetry
in their distributions.\\
$\bullet$ out of equilibrium dynamics---the Universe
is often an almost-thermalised bath, and in 
chemical equilibrium, there are no
asymmetries in  unconserved quantum
numbers (such as $B$, by the first condition).\\

These ingredients, including baryon number violation,
  are all present in the Standard Model. 
The SM contains C and CP violation,  parametrised by
the Jarlskog invariant  $\sim 10^{-23}$. To use such a small
amount of CP violation to generate $Y_B \sim 10^{-10}$
is not obvious; maybe it could arise via some kinematic
amplification  factor, such  as
gives $\epsilon \sim 10^{-3}$
 in  $\bar{K} -K$ oscillations. However, this
is difficult in  a thermal bath \cite{Gavela:1994dt}.

The non-equilibrium can arise due to the
expansion of the Universe, for instance, from an interaction
whose timescale is of order the age of the Universe.
This happens in thermal leptogenesis, where
the asymmetry generating interactions of the right-handed neutrino
occur on the timescale  Hubble$^{-1}$.

The third ingredient, Baryon number violation,
is (unexpectedly) present and fast \cite{Rubakov:1996vz} in  SM cosmology.
Non-perturbative knots in the SU(2)  gauge fields can act as sources for
$B+L$, emitting simultaneously  one lepton and three quarks  
of each generation \footnote{Kinematically, these cannot
mediate proton decay. Also, at zero temperature
the knots are instantons, with an exponetially
suppressed rate $\sim e^{-8 \pi/g^2}$ \cite{'tHooft:1976up}}. 
 The rate for tying and untying these knots,
somewhat abusively refered to as sphalerons \cite{Klinkhamer:1984di},
is  fast  before  the electroweak phase transition. 
This will be the source of baryon number violation
used in leptogenesis: a lepton asymmetry is produced 
by some mechanism, then  the fast SM $B+L$
eating interactions partially reprocess it
\cite{Khlebnikov:1988sr} to a baryon asymmetry.

Although the ingredients for baryogenesis
are all present in the SM,  a way to combine them
to generate the observed baryon asymmetry has not been found.
The baryon asymmetry 
 is therefore taken as evidence for Beyond the SM  (BSM) 
physics.

\subsection{Why leptogenesis?}
 
It is  interesting to
generate the baryon asymmetry in BSM models that are motivated for
other reasons, because
BSM models usually have many  free parameters,
and $Y_B$ is just one number.

One of the challenges in building baryogenesis models
is the non-observation of proton decay. So it is
convenient to  use for baryogenesis  B violation
that does not cause the proton to decay--- such as
the SM non-perturbative processes.

Neutrinos are observed to have small masses, which 
could be Majorana, and therefore lepton number violating.
The same lepton number violation that generates
neutrino masses could be used to generate a cosmological
lepton asymmetry, transformed to a baryon asymmetry
by the SM non-perturbative processes.

Leptogenesis in the seesaw is therefore
an attractive possibility for baryogenesis,
because  the proton can be stable, and 
the seesaaw is a popular neutrino mass generation mechanism.

\section{The Seesaw and Leptogenesis}

\subsection{The model}

The seesaw model\cite{seesaw}  naturally explains
the small observed neutrino masses. In
its most simple formulation,   two or
three  right-handed neutrinos $N_i$,
 are added the Standard Model.
Being gauge singlets,  they can
have  large Majorana masses, and
the Lagrangian can be written in
the mass basis of the charged leptons and
right-handed neutrinos \footnote{The Yukawa indices 
are ordered left-right,  and 
the $*$ on $\lambda$  reproduces
the Lagrangian of the superpotential
$W = LH   \lambda N^c + N^c \frac{M}{2} N^c $} 
 as:
\bea
 {\cal L}& =& {\cal L}_{SM} - 
[{\bf \lambda}]^*_{\alpha k} {\overline{\ell}}_\alpha  { N_k 
\cdot \phi}
- \frac{1}{2}{\overline{N_{j}}}{M_j}{N_{j}^c}
\label{L}
\eea
There are then
21 parameters in the  lepton sector of the Lagrangian
(\ref{L}): counting in
the $N$ and charged lepton mass eigenstate bases, there
are the six masses $m_e, m_\mu, m_\tau,$ $   M_1, M_2, M_3$  and 
18 - 3 phases, angles and eigenvalues in $\lambda$
(three  unphysical phases can be removed by judicious choice of $\ell$
phases).

At scales $\ll$ $M_1$, this gives an effective light
neutrino mass matrix
\bea
[m_\nu] &  =  &   {\bf  \lambda}
{ M}^{-1} {\bf  \lambda}^T v_u^2 ~~~~~~~
 ~ (v_u = \langle \phi^0\rangle \simeq   175 ~{\rm  GeV})
\label{mass}
\eea
In the leptonic sector of the  SM  augmented with
the Majorana neutrino mass matrix of eqn (\ref{mass}), there are
12 parameters: 
$m_e, m_\mu, m_\tau,$ the neutrinos masses 
$ m_1, m_2, m_3$  
 and 3 angles and  3 phases  in the mixing matrix
 $U_{PMNS}$ between the eigenbases. Seven of these
parameters are measured,  there is an upper bound on
the mixing angle $\theta_{13}$, and the light neutrino mass
scale and three phases of $U_{PMNS}$  are unknown.
  There are in addition
9 unknown parameters in the high scale theory, which
hopefully arrange themselves such that leptogenesis can work.

\subsection{Leptogenesis Mechanisms}

In the context of the seesaw extension of the SM, there
are many ways to produce a baryon asymmetry. They differ
in the cosmological scenario, and  in the values of undetermined
seesaw parameters. An incomplete list of  possibilities is:
\begin{itemize}
\item   ``Thermal'' leptogenesis with hierarchical $M_j$
\cite{Fukugita:1986hr,Buchmuller:2004nz,Giudice:2003jh},
 will be discussed in the next sections.
The $N_1$ are produced by scattering in the thermal bath.
\item Thermal leptogenesis with quasi-degenerate $M_j$
\cite{Pilaftsis:2005rv}  can work
for lower reheat temperatures, because the $CP$ violation
can be enhanced in $N_i-N_j$ mixing.
\item ``soft leptogenesis''\cite{Grossman:2004dz}  
can work  in a  one-generational SUSY seesaw.
If the soft SUSY-breaking terms  are of suitable size,
there is  enough  $\CPv$ in  $\tilde{N} -\tilde{N}^*$ mixing.
\item In the  Affleck Dine mechanism \cite{Dine:1995kz},
an asymmetry arises in a classical scalar field,
which later decays to particles.  The field
starts with a large expectation value, which gives it access
to lepton number violation that is suppressed at small scales.
\item The  $N$ could be produced
non-thermally, for instance in inflaton decay
\cite{Asaka:1999yd}, or  in preheating\cite{Boubekeur:2002gv}.
\end{itemize}

\section{Thermal Leptogenesis}

The remainder of this proceedings focusses on
the first scenario  (thermal production of 
hierarchical $N$). It  assumes
\begin{enumerate}
\item the Lagrangian  of eqn (\ref{L})
\item hierarchical $N$ masses: $M_1 \sim 10^{9} ~{\rm GeV} \ll M_2, M_3$
(a hierarchical spectrum  seems indicated, if $\lambda$  is hierarchical.
Also, the kinematics is  simpler in an effective theory
of propagating $N_1$ and effective dimension 5 operator
induced by $N_2$ and $N_3$).
\item  thermal production of
the $N_1$ (and negligeable production of $N_2$) 
\footnote{If the reheat temperature of the Universe
is $> M_2$, then the asymmetry produced in $N_2$
decay can be relevant \cite{Vives:2005ra,EGNN,barbieri}}.
\end{enumerate}

The idea is that a distribution of $N_1$ is
produced by scattering processes at temperatures
$T \sim M_1$, and then the $N_1$ decay away, 
as the temperature drops below their mass,
because the equilibrium number density is
 suppresssed $\propto e^{-M_1/T}$. If these decays are
CP violating,  asymmetries in all the lepton flavours can be produced. 
If inverse decays are out-of-equilibrium, the
asymmetries may survive. 
They can then be reprocessed into a baryon asymmetry by
the SM B+L violating processes.

The baryon asymmetry can be estimated by
considering  the Sakharov conditions. The maximum baryon 
asymmetry would arise if, 

\begin{enumerate}
\item
at $T \gg M_1$ there was a thermal
distribution of $N_1$, 
\item each $N_1$  contributes, when it 
 decays,
1 lepton to the  asymmetry in some flavour $\alpha$,
\item   there are no inverse decays to wash out
the asymmetry produced,  
\item  each lepton then
is converted into a baryon.
\end{enumerate}
 More realistically: 
\beq
Y_B = \frac{n_B - n_{\bar{B}}}{s}
\simeq
\frac{135 \zeta(3)}{4 \pi^4g_*}\times
(prefactor) \times  \sum_{\alpha} \epsilon_{\alpha \alpha}  \times \eta_\alpha 
\label{approx}
\eeq
where 
the first  fraction is  the equilibrium
$N_1$ number density divided by the  entropy density at
$T \gg M_1$,   of order $ 4 \times 10^{-3}$ when the number of relativistic
degrees of freedom  $g_*$ is taken $ \simeq 106$
as  in the SM.  The non-equilibrium
is parametrised in  $\eta_\alpha $, which varies
from 1 (fully out-of-equilibrium) to 0 (decay
in equilibrium, no asymmetry).
 The  asymmetry in lepton flavour $\alpha$  per
$N_1$ decay  is $ \epsilon_{\alpha \alpha} \ll 1 $ (double
index because formally it is a diagonal element of a matrix), 
and the prefactor $ \sim 1$  converts the lepton asymmetry
produced in $N_1$ decay  to a baryon asymmetry. 
The aim is now  to  estimate these  parameters.

The estimates will be performed for a lepton flavour
$\alpha$. At the end, we can sum over flavour.


\subsection{L and B+L violation}

The interactions of $N_1$ violate L  because
lepton number cannot be consistently assigned to
$N_1$ in the presence of $\lambda$ and $M$: 
if $N_1$ is a lepton, then $M$ violates L by two
units, if $N$ is not a lepton, then
$\lambda$ violates L by one unit.  
The $N_1$ decay, which depends on $M_1$  and
$\lambda$,  does not conserve L: being  its own anti-particle,
$N_1$ can decay  to either  $\ell \phi$ or $\bar{\ell} \phi^*$. 
If there is an asymmetry in the rates, a net
lepton asymmetry will be produced.

The baryon number violation is provided by
 $B+L$  changing SM non-perturbative processes, which
are fast \cite{Rubakov:1996vz}   
(The rate $\Gamma_{B-L} \sim \alpha^5 T$ is faster than the
Hubble expansion $ H$) in the thermal plasma 
below $T \lsim 10^{12} $ GeV. 
An asymmetry in lepton flavour $\alpha$,
 produced in the $N_1$ decay,
contributes to the density of   $B/3-L_\alpha$, which
in conserved  by the SM interactions. In equilibrium,
this excess of  ${B-L}$ implies (for the SM)  
\cite{Khlebnikov:1988sr} a baryon
excess
\beq
Y_B \simeq  \frac{12}{37} \sum_\alpha Y_{B/3-L_\alpha}
\label{prefactor}
\eeq
12/37 is the prefactor in the SM.

\subsection{CP in the decay }

To produce a net lepton  asymmetry,  the  $N_1$  must 
have different  decay
rates  to final states
with particles or anti-particles. 
The  asymmetry in lepton flavour $\alpha$,
produced in the decay of $N_1$ is
\bea
 \!  \! \! \epsilon_{\alpha \alpha} & =& \frac{    \!
 \Gamma(N_1  \! \rightarrow 
\!  H \ell_\alpha)
- \Gamma(\bar{N}_1  \!\rightarrow  \! \bar{H} \bar{\ell}_\alpha) \!}{
\Gamma(N_1  \! \rightarrow  \! H \ell) \! +  \!
 \Gamma(\bar{N}_1 \rightarrow  \! \bar{H} \bar{\ell})}
\eea
The $\epsilon_{\alpha \alpha}$ carries an incongruous
double flavour index, to remind us that it is
the diagonal element of a matrix, and not  a
component of a vector (this is relevant for
writing Boltzmann Equations). It is normalised to
the total decay rate, so that  the  Boltzmann
Equations are linear in flavour space.

\begin{figure}[ht]
\unitlength0.5mm
\SetScale{1.4}
\begin{boldmath}
%
\hspace{1cm}
\begin{picture}(60,40)(0,0)
\Line(0,0)(30,0)
\Text(15,0)[c]{X}
\ArrowLine(30,0)(60,0)
\DashLine(30,0)(60,30){1}
\Text(30,-5)[c]{{$\lambda$}}
\Text(-2,0)[r]{{$N_1$}}
\Text(57,-6)[l]{{$\ell_\alpha$}}
\Text(62,30)[l]{{$\phi$}}
\Text(63,0)[l]{{$~~\times {\Big (}$}}
\end{picture}
\hspace{1cm}
\begin{picture}(60,40)(0,0)
\Line(15,0)(0,0)
\DashLine(45,0)(15,0){1}
\ArrowLine(45,0)(60,0)
\ArrowArc(30,0)(15,0,180)
\DashLine(40.6,10.6)(60,30){1}
\Text(15,-10)[c]{{$\lambda^*$}}
\Text(45,-10)[c]{{$\lambda$}}
\Text(40,23)[c]{{$\lambda$}}
\Text(55,7)[c]{{$N_{2,3}$}}
\Text(44,4)[c]{X}
\Text(27,-5)[l]{{$\phi$}}
\Text(17,17)[c]{{$\ell$}}
\end{picture}
\hspace{1cm}
\begin{picture}(60,40)(0,0)
\Text(-10,0)[r]{{$+$}}
\Line(15,0)(8,0)
\DashLine(45,0)(15,0){1}
\Line(45,0)(60,0)
\ArrowLine(60,0)(80,0)
\ArrowArc(30,0)(15,0,180)
\DashLine(60,0)(80,20){1}
\Text(15,-7)[c]{{$\lambda^*$}}
\Text(45,-7)[c]{{$\lambda$}}
\Text(62,-7)[c]{{$\lambda$}}
\Text(51,10)[c]{{$N_{2,3}$}}
\Text(52,0)[c]{X}
\Text(30,-5)[l]{{$\phi$}}
\Text(20,18)[c]{{$\ell$}}
\Text(85,0)[c]{{$~{ \Big ) }$}}
\end{picture}
\vspace{.3cm}
\end{boldmath}
\label{fig1}
\caption{The diagrams contributing to  the
CP asymmetry $\epsilon_{\alpha \alpha}$. The
flavour of the internal lepton $\ell$ is summed,
and the internal $\ell$ and Higgs $\phi$ are on-shell.
The $X$ represents a Majorana mass insertion.}
\end{figure}
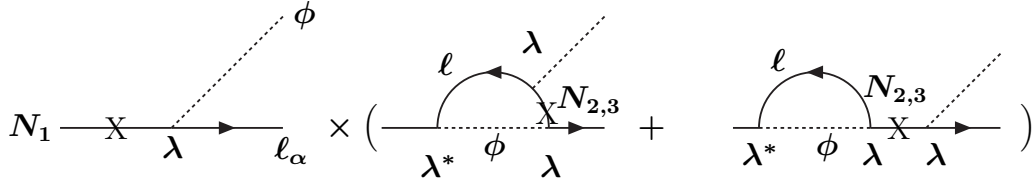
The  CP asymmetry $\epsilon_{\alpha \alpha}$
arises  from the  intereference of the
tree level and one-loop amplitudes.
Writing the tree + loop  amplitude as a product of coupling
constants $c$ and everything else: 
$c^t {\cal A}^t + c^{loop} {\cal A}^{loop}$,
one finds that the CP asymmetry is proportional
to $\Im \{ c^{t*} c^{loop} \} \times \Im \{ {\cal A}^{t*}{\cal A}^{loop} 
\}$ (where the imaginary part of
the  amplitudes arises form putting the loop particles  on-shell).  
This is straightforward
to calculate from the diagrams in figure 1
\cite{Covi:1996wh}.  For hierarchical right-handed neutrinos
($M_{2,3} \gg 10M_1$), the internal $N$ propagator can
be collapsed to the effective operator   $[m_\nu]/v_u^2$
(see eqn (\ref{mass}) \footnote{This approximation would
not work if $N_1$ propagated in the loop; it does not because
the  associated coupling constant combination is real.}), which gives
\beq
\epsilon_{\alpha \alpha} = \frac{3 M_1}
{16 \pi v_u^2 [\lambda^\dagger \lambda]_{11}} 
\Im \{ [\lambda]_{\alpha 1} [m_\nu^* 
\lambda]_{\alpha 1} \} 
\label{epsaa}
\eeq
The total CP asymmetry is bounded above \cite{Davidson:2002qv,HMY}
\beq
\sum_\alpha \epsilon_{\alpha \alpha} < 
\frac{3}{16 \pi} \frac{(m_3 - m_1) M_1}{<\phi>^2} 
\sim 10^{-6}  \frac{M_1}{10^{9} {\rm GeV}}
\label{di2}
\eeq
This suggests a lower  bound  on the mass of $N_1$, and
the reheat temperature,  of order 
 $ M_1 \gsim 10^9 ~{\rm GeV}$. This applies 
only to hierarchical $N_i$; the  CP asymmetry
can be much larger for quasi-degenerate $N_i$.

\subsection{Out of Thermal Equilibrium}

The non-equilibrium for thermal leptogenesis is
provided by the Universe expansion: interaction rates which
are of order the Hubble expansion rate $H$, are not fast
enough to equilibrate particle distributions. This is
the most delicate part of estimating the baryon asymmetry.

Suppose, as initial conditions, that after inflation
the Universe reheats to  a thermal  bath  composed of particles with
gauge interactions.  The $N_1$ can be produced by
inverse decays $\phi \ell_\alpha \rightarrow  N$, and
most effectively by scattering $  q_L t_R \rightarrow 
\phi \rightarrow \ell_\alpha N$.
A thermal number  density of $N_1$ ($n_N \simeq n_\gamma$)   
will be produced if $ M_{1}  \lappeq T$, and  if
 the production timescale
 for $N_1$s, $1/\Gamma_{prod}$,
 is  shorter than the age of the Universe $ \sim 1/H$:
\beq
\Gamma_{prod} \sim  \sum_\alpha \frac{h_t^2 |\lambda_{\alpha 1}|^2}{4 \pi} T > H 
\eeq
If this is satisfied, then, since $h_t \sim 1$,
  the  $N_1$ decay  is also ``in equilibrium'':
\beq
 \Gamma_{D} \simeq  \frac{ [\lambda^\dagger \lambda]_{11} M_1}{8 \pi}
> \left. \frac{10 T^2}{m_{pl}}\right|_{T = M_1}
\eeq 
It is possible to show that
\beq
\tilde{m} = \frac{[\lambda^\dagger \lambda]_{11}v_u^2}{M_1}
\label{tildem}
\eeq
 the total decay rate  rescaled by factors of $M_1$ and
Higgs vev, 
 is ``usually''  $ \gsim  m_{sol}$, so one
expects a maximal initial distribution of
$N_1$, and a  total decay rate that is fast compared  to $H$.

More importantly,   as the $N_1$ start to decay,
 the inverse decays
$\ell_\alpha \phi \rightarrow N_1$, which can
wash out the asymmetry, 
may be fast compared to $H$.
Suppose this is the case for flavour $\alpha$. 
 Then the asymmetry  in lepton  flavour
$\alpha$    will
survive  once  inverse decays from flavour
$\alpha$  are ``out of equilibrium'':
\beq
\Gamma_{ID} (\phi \ell_\alpha \rightarrow N_1) \simeq 
\Gamma_{\alpha \alpha} e^{-M_1/T}
<  \frac{10 T^2}{m_{pl}}   
\label{10}
\eeq
where  $\Gamma_{\alpha \alpha}  \equiv
\tilde{m}_{\alpha \alpha}M_1^2/(8 \pi v_u^2) $ is the partial decay
rate  $\Gamma(N \rightarrow \ell_\alpha \phi)$. 

At  temperature  $T_\alpha$  where eqn (\ref{10}) is satisfied,  
the remaining
$N_1$ density    is Boltzmann suppressed: $ \propto   
e^{-M/T_\alpha}$.
Below $T_\alpha$,  the $N_1$  decay `` out of equilibrium'',
and  contribute to the lepton flavour asymmetry.  So the washout factor
$\eta_\alpha$ for flavour $\alpha$ can be approximated
\footnote{There is a subtlety: the asymmetry  produced in $B/3 - L_\alpha$
 is spread amoung different particle species by
the SM interactions (for instance,
some of  the $\ell_\tau$ asymmetry could be
stored in $\tau_R$). But washout proceeds
only from the asymmetry in doublets, so is
mildly reduced, depending on which SM interactions
are fast enough  to redistribute the asymmetry.
As discussed in \cite{matters,AAFX}, this can 
usually by accounted for using
the $A$-matrix of \cite{barbieri}.} as
\beq
\eta_\alpha \simeq \frac{ n_N(T_\alpha)}{ n_N(T\gg M_1)} \simeq e^{-M/T_\alpha}
 \simeq \frac{m_*}{\tilde{m}_{\alpha \alpha}}
\label{eta}
\eeq
where ${m_*}/{\tilde{m}_{\alpha \alpha}}$ 
is ${H}/{\Gamma_{\alpha \alpha}}$  evaluated at $T = M_1$.  This approximation
applies for $\tilde{m}_{\alpha \alpha} > m_* \simeq
10^{-3}$eV.

\section{With or without flavour?}


Consider the case of strong washout for all flavours.
 Combining equations  (\ref{approx}), (\ref{prefactor}),
(\ref{eta}) and (\ref{epsaa}) gives
\beq
Y_B  \sim 10^{-3} \, { \sum_\alpha} \epsilon_{\alpha \alpha} \eta_\alpha
\sim  10^{-3}  \, m_*  \, { \sum_\alpha} 
\frac{ \epsilon_{ \alpha \alpha} } {
 \tilde{m}_{\alpha \alpha}}~~~ ~~~{\rm (flavoured,~ strong~ washout)}
\label{matter}
\eeq
where the  flavours summed over are  presumably 
the charged lepton mass eigenstates.  However,
one might also think to  choose  $ \alpha$ as
the direction in flavour space into which
$N_1$ decays, which can be called $\hat{y}$.
 Then  $ \epsilon_{yy}$ is the total
CP asymmetry \footnote{This is technically not quite correct,
 because the final lepton states are not CP eigenstates.
See \cite{Nardi:2006fx}.} $ \epsilon$, $\Gamma(N \rightarrow \phi \ell_y)$
is the total decay rate $\Gamma_D$, and one finds
\beq
Y_B \sim   10^{-3} \epsilon   \, \frac{ m_* }{\tilde{m}}
~ ~~~~~~({\rm single ~ flavour, strong~ washout})
\label{um}
\eeq
Which is simpler but  not the same.
 The remainder of this  section discusses why and
when the charged lepton mass eigenstate basis
is the relevant one.

\subsection{But flavour should be irrelevant...?}

Flavour  effects were ignored in leptogenesis for
a long time. One reason  is that
 the small
CP asymmetry in $N_1$ decay depends on many
powers of  $\lambda$. 
Including the  charged lepton Yukawas  should
be a perturbatively irrelevant correction.
Secondly,  the $B+L$ violating processes
partially transform the {\it total} lepton asymmetry
to baryons, and the total lepton asymmetry is
the trace of an ``asymmetry number  operator'' 
 in flavour space. 
Since the trace can be evaluated in any basis,
the simplest  choice  would  include  the direction
$\hat{y}$ into which $N_1$ decays.

Nonetheless, flavour matters\cite{issues,Nardi:2006fx,matters}
: equations
(\ref{matter}) and (\ref{um}) are different, 
because the first   sums probabilities and
the second sums amplitudes.
Thermal leptogenesis
always  involves the production and decay
of an $N$, via its Yukawa vertex where
also appears  the lepton $\ell_y$ 
\footnote{For instance,
in the strong washout example of the previous
section,  the $N_1$  number density is
depleted by decays and repopulated by inverse decays.
A lepton doublet is involved in both of
these interactions.}.  If lepton
flavours are indistinguishable between 
these two interactions, then the ``single
flavour'' results are correct. But if
the charged lepton Yukawas give distinguishable
thermal masses to  different flavours,
then the asymmetry should be computed
flavour by flavour.

The   timescale for leptogenesis is $H^{-1}$ (because
the ``non-equilibrium'' is provided by the Universe
expansion).  One way to take into
account the many  interactions that are  faster than $H$
is to resum them  into thermal masses.
 Comparing $H$  to the rates for   $h_\tau$ or  $h_\mu$ 
mediated interactions such as $q_L \overline{t_R} 
\rightarrow \ell_\tau \overline{\tau_R}$), one finds
\bea
\Gamma_\tau \simeq 10^{-2} h_\tau^2 T  > H {\rm ~  for ~} T < 10^{12} ~ GeV,
 ~~~~\Gamma_\mu > H~ {\rm  for}~ T < 10^9  ~ GeV \nonumber
\eea
So below $T \sim 10^{12} $ GeV,
 the $h_\tau$ is ``in  equilibrium'',  
and contributes to the ``thermal mass  matrix''
\footnote{ If $\Gamma(N \rightarrow \phi \ell) > \Gamma_\tau$, it should also
be included in determination of mass eigenstates,
as noted in \cite{Blanchet:2006ch}} of the lepton doublets.
So there can be two distinguishable flavours down to
$T \sim 10^{9} $ GeV, and below $T \sim 10^{9} $ GeV
there can be three. In the scenario discussed here, where
the asymmetry is generated in the decay of $N_1$,
the temperature of leptogenesis $\gsim 10^9$ GeV,
to obtain a large enough CP asymmetry (see eqn (\ref{di2})).

The second reason given above, for why flavour is
`` obviously'' irrelevant, is elegant and convincing.
 A technical way to see what is missing  is to
write  the equations  of motion for   the asymmetry
number operator, 
which is a matrix in flavour space (see  \cite{barbieri},
or the appendix of \cite{issues} for a toy model).  The diagonal
elements of this  matrix are the asymmetries, the
off-diagonals encode quantum correlations (as
is the case for a quantum mechanical density matrix). 
These equations transform  under  changes of the flavour basis, and
contain terms describing the interactions mediated by
the charged lepton Yukawa matrix $[h_e]$ . In the charged lepton
mass eigenstate basis, the $[h_e]$  terms affect 
only the off-diagonals, and  drop out of the
Boltzmann-like equation for the flavoured lepton
asymmetries on the diagonal. But in any other basis,
these $[h_e]$  terms should remain.

\section{In practise, so what?}

As discussed after eqn (\ref{mass}), there are (currently)
14 unknown parameters in the seesaw model, of which
only a few are accessible to low energy experiments.
It is interesting, from a phenomenological perspective, to constrain
the unknowns by requiring that leptogenesis works. 
Including flavour effects in calculating
these   constraints does not change 
them very much, as discussed in  the following subsection.

From a top-down  perspective, models give predictions
for all the parameters of the seesaw. It is interesting
to verify that a given model reproduces   the correct baryon
asymmetry, as well as the observed neutrino masses. 
Including flavour in such  leptogenesis
calculations can make a significant difference. 
This is discussed  later.

\subsection{Phenomenological Bottom-up perspective}
\label{pheno}

\begin{enumerate}
\item { The  upper  bound on the light neutrino mass scale:}\\
In the ``single flavour'' calculation, successful thermal leptogenesis
required a  light $\nu$ mass scale $ \lsim .2$ eV 
\cite{Buchmuller:2003gz}. This is
no longer the case in the flavoured calculation---
models can be tuned to work for  $m_\nu \lsim $ few eV ( the
cosmological bound).  This is because  there is  more $\CPv$
available.  The upper limit of eqn
(\ref{di2})  on  the total CP asymmetry, {\it decreases}
like $\Delta m_{atm}^2/m_{max}$, as the light neutrino
mass scale  $m_{max}$ increases. There is
therefore an upper bound on $m_{max}$.  
However, the limit (\ref{di2}) does not
apply to the flavoured CP asymmetries, which can 
increase with the light neutrino mass scale.

\item { Sensitivity of the baryon asymmetry to 
PMNS phases:}\\
An important, but sad, observation in
``single-flavour'' leptogenesis  was 
the lack of a  model-independent  connection between
$\CPv$ for leptogenesis and MNS phases. It was shown
\cite{Branco:2001pq} that thermal leptogenesis can work
with no $\CPv$  in $U_{PMNS}$, and conversely,
that leptogenesis can fail in spite
 of phases in $U_{PMNS}$.   In the
  ``flavoured'' leptogenesis case,
it is still true that the   baryon asymmetry  is
not sensitive  to PMNS phases \cite{todo} 
(=leptogenesis can work  for any value of
the PMNS phases). However, interesting observations
can be made in classes of models \cite{Nardi:2006fx,Pascoli:2006ie,Branco:2006ce}.

\item { The lower bound on $T_{reheat}$:}\\
There is an envelope, in  the space of parameters leptogenesis
depends on,  
inside which  leptogenesis {\it can} work.
In the ``single-flavour'' calculation, the
most important parameters are $M_1$, $\Gamma$ (equivalently
$\tilde{m}$), $\epsilon$ and
the light neutrino mass scale \cite{Buchmuller:2004nz}. 
 Including  flavour gives the  envelope more dimensions
($M_1, \epsilon_{\alpha \alpha}, \Gamma_{\alpha \alpha}$...),
but it can still be  projected onto $M_1$, $\tilde{m}$ space.
Leptogenesis works for $M_1$ a factor
of $ \sim 3 $ smaller in the
 ``interesting'' region of $m_* < \tilde{m} \lsim m_{atm}$.
But the lower bound on $M_1$, in the
 $m_* \sim  \tilde{m} $ region \footnote{A smaller $M_1$ could be possible for
very degenerate light neutrinos\cite{issues}},
remains $\sim 10^{9}$ GeV 
\cite{Blanchet:2006be,Antusch:2006gy,AAFX}.

\end{enumerate}

\vspace{2cm}

\subsection{Estimating the Baryon Asymmetry in Models}
\label{topdown}

In the ``single flavour''  calculation, the baryon
asymmetry can be approximated \cite{Giudice:2003jh} as 
\beq
Y_B \simeq  4 \times10^{-3}  \epsilon \eta ~~~~~~~~~~
 \eta \simeq  \frac{1}{\tilde{m}/m_* + m_*/5\tilde{m}} 
\label{without}
\eeq
where $\tilde{m}$ is the rescaled decay rate (see eqn (\ref{tildem}),
and $m_* \simeq 10^{-3}$eV is the  value of $\tilde{m}$ for which $\Gamma_D = H$
at $T = M_1$.
So to estimate the baryon asymmetry produced by leptogenesis
in a particular model, one merely must calculate $\epsilon$
and $\tilde{m}$.

In the flavoured case, one should
work two or three times harder: if 
the tau  Yukawa is  ``in equilibrium''
(and $h_\mu$ not), there are two relevant  
CP asymmetries
$\epsilon^{ee} + \epsilon^{\mu \mu},
\epsilon^{\tau \tau} $, and two  partial
decay rates  $\tilde{m}_{ee} + \tilde{m}_{\mu \mu},
\tilde{m}_{\tau \tau} $. 
Assuming that\footnote{This means  at least one flavour 
is in strong washout. 
 An approximate formula for $Y_B$ in the case of
$\tilde{m} < m_*$ can be found in \cite{matters}.}
  $\tilde{m} > m_*$, the  final baryon asymmetry
can be approximated \cite{matters}
 as
\beq
 Y_B \simeq     ~~~ 4 \times 10^{-3} \sum_\alpha 
\epsilon_{\alpha \alpha}  \eta^{ \alpha}~~~~~~~~~~~~~
 \eta^{\alpha } \simeq \frac{1}{\tilde{m}^{\alpha \alpha} /m_* + m_*/5\tilde{m}^{\alpha \alpha}}
\label{with}
\eeq

\begin{figure}[h]
\centerline{\hspace{-1cm}
\epsfxsize=7cm\epsfbox{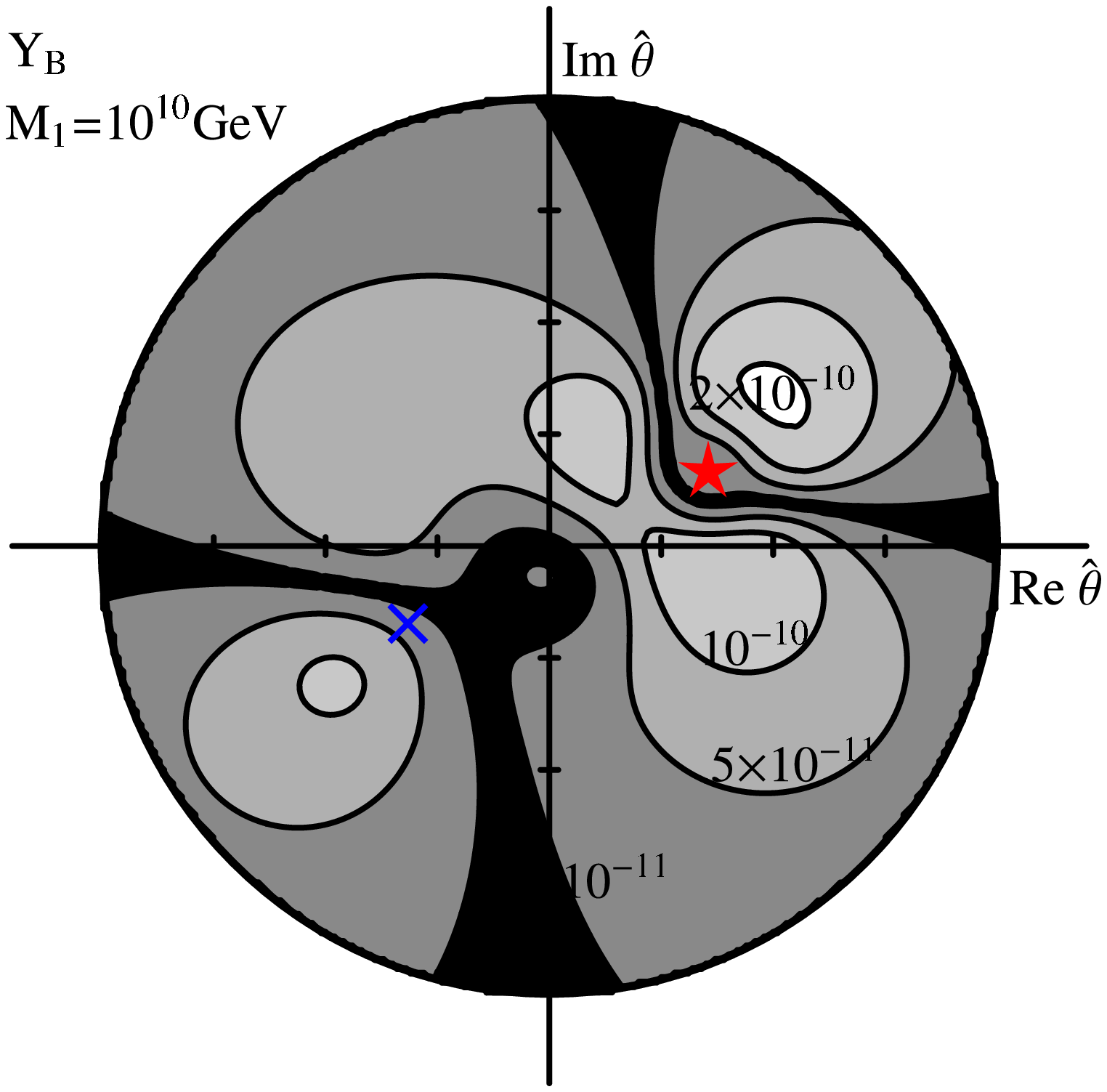}
\hspace{-0.5cm}
\epsfxsize=7cm\epsfbox{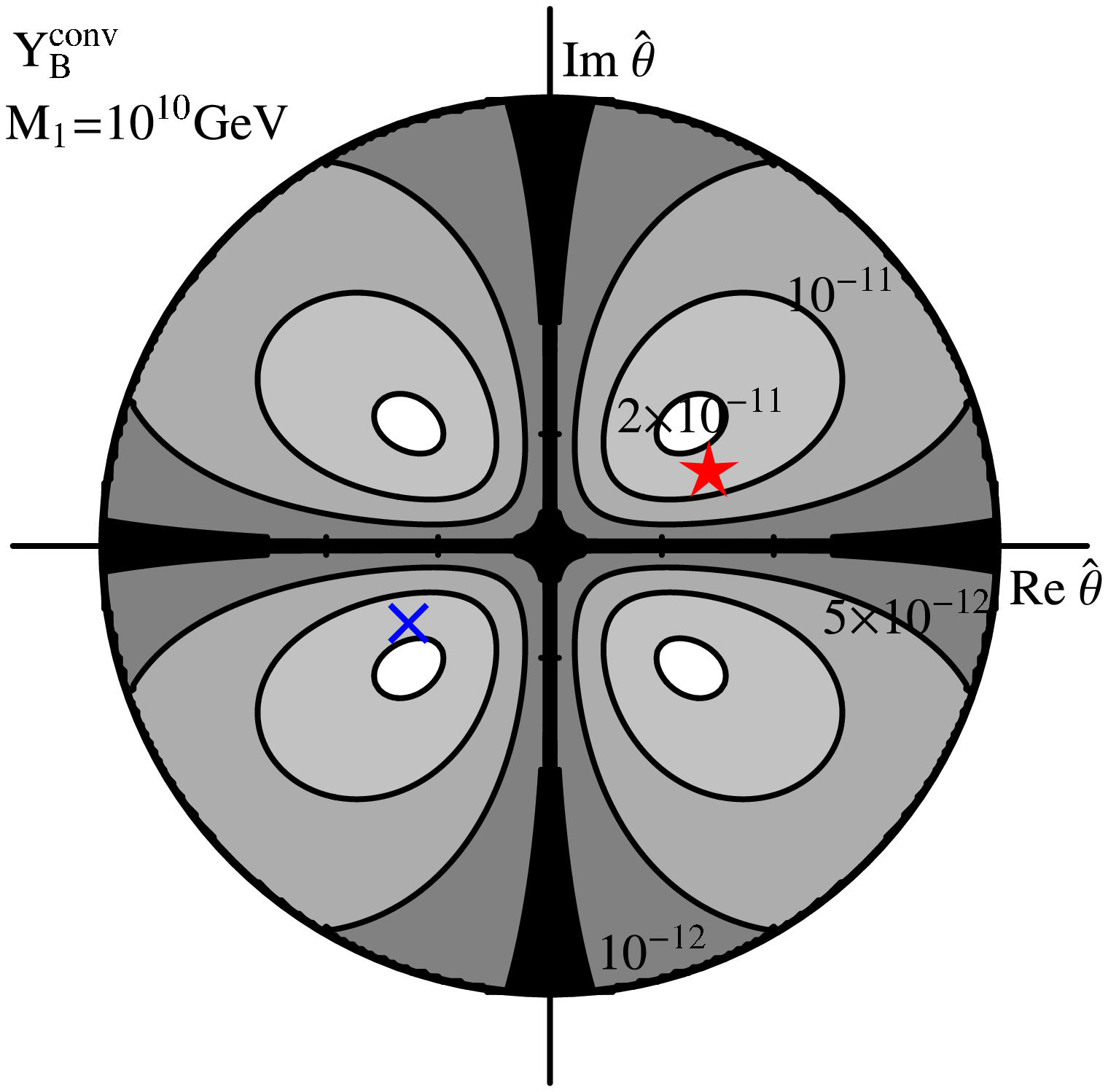}}
\caption{ Contour plots of 
$Y_B$ in the 2 right-handed neutrino  model,  with 
hierarchical $\nu_L$.On the
LHS  flavoured leptogenesis, the  RHS  is the ``single flavour'' result. 
In the flavoured plot, the white region corresponds
to a large enough asymmetry. In the unflavoured
calculation, the white regions are an order of magnitude smaller.
The blue cross [red star]  correspond to a (1,2) texture zero
[(1,3) texture zero].
The plot is in the  plane of the one  
complex angle of the  R matrix required to parametrise 
this model. See \protect\cite{matters} for details.}
\protect\label{ibarra}
\end{figure}

The flavoured formula (\ref{with}) can give a significantly
larger  result than eqn (\ref{without}), as can be seen
in figure \ref{ibarra}. There are two intuitive reasons
for this. First, the washout of the asymmetry
by  inverse decays is less efficient with flavour,
because inverse decays from flavour $\alpha$ only
can  destroy the asymmetry in flavour $\alpha$. As opposed to the
total inverse decay rate  eating  the whole lepton
asymmetry, as in the single flavour case.
Second, including the charged lepton Yukawas
puts more $\CPv$ in the theory. So the flavoured
CP asymmetries can be larger than the sum, 
because this additional CP violation must vanish
from $\epsilon$. 

Finally, consider the case where
the light neutrino masses are
non-degenerate, and 
 all ({\it e.g.}, two) 
distinguishable flavours  
are in strong washout. Then there is a pretty  approximation
to $Y_B$ \cite{todo},  which exhibits the amplification of
the asymmetry due to flavour effects. The
asymmetry  can be approximated
\beq
Y_B \simeq  \frac{12}{37}\frac{135 \zeta(3)}{4 \pi^4 g_*}
\sum_\alpha \epsilon_{\alpha \alpha} \frac{m_*}{5 \tilde{m}_{\alpha \alpha}}
\label{18}
\eeq 
and the ratios in the flavour sum can be expressed
\beq
\label{19}
m_* \frac{ \epsilon_{\alpha \alpha}}{ \widetilde{m}_{\alpha \alpha }}=
\frac{3 M_1m_*}{16 \pi  \tilde{m}} \sum_\beta 
\Im \{ p_\alpha  [m_\nu]^*_{\alpha \beta} 
p_{\beta} \} \left| \frac{\lambda_{\beta 1}}{ \lambda_{\alpha 1}} \right|
~~~~~~~~
{\rm where}~ {p}_\alpha \equiv \frac{\lambda_{\alpha 1}}{  |\lambda_{\alpha 1 }|} 
\eeq
(no sum on $\alpha$).
Substituting eqn (\ref{19}) into eqn (\ref{18}) gives
that the  flavoured asymmetry, divided by
the unflavoured upper bound  (see eqns (\ref{um}),(\ref{di2})), 
will be \cite{todo}
\beq
 \simeq \left(   \frac{\Im \{ p_\tau  [m_\nu]^*_{\tau \tau} 
p_{\tau} \} }{m_{atm}} +
 \frac{\Im \{ p_o  [m_\nu]^*_{oo} 
p_{o} \} }{m_{atm}} + 
 \frac{\Im \{ p_\tau  [m_\nu]^*_{\tau o} 
p_{o} \} }{m_{atm}}\left[ \frac{|\lambda_{o1}|}{|\lambda_{\tau 1}|} +
 \frac{|\lambda_{\tau 1}|}{|\lambda_{o 1}|} \right] \right)
\label{belle}
\eeq
Recall that this equation is only valid in strong washout
for all flavours, and that the $p_\sigma$ are the  phases
of the Yukawa couplings
\footnote{In general, these phases are related to
the light neutino mass matrix, so should not be chosen
independently.}. The flavour $o$ is the projection of
$\hat{y}$ (defined after eqn (\ref{matter})) on $e, \mu$ space.
The bracketed term  shows how  stronger
washout in one flavour  can increase  the baryon asymmetry. 
So models in which the Yukawa coupling $[\lambda]_{\tau 1}$
is significantly different from $[\lambda]_{\mu 1},[\lambda]_{e 1} $,
can have an enhanced baryon asymmetry (with
cooperation from the phases).

This equation is attractive  step towards
writing the baryon asymmetry as a real function
of real parameters (the unflavoured
upper bound on $Y_B$, depending on  $M_1$ and $\tilde{m}_1$),
times a phase factor \cite{HMY}. In this case,
the phase factor is a sum of three terms, depending on:
the phases of the neutrino Yukawa couplings,  light
neutrino mass matrix elements normalised by the heaviest mass,
and a real ratio of Yukawas.

\section{Summary}

Thermal leptogenesis  in the seesaw model is an attractive
baryogenesis  mechanism. Some
density of  right-handed
neutrinos $N$ is generated by scattering in the plasma, then
the $N$ produce a lepton asymmetry in their decay.
Standard Model $B+L$ violating processes partially
transform this lepton asymmetry to  baryons. The  right-handed
neutrino masses  were taken  hierarchical  in this proceedings.

When  the interaction rates of the charged lepton Yukawas are 
faster than the leptogenesis rates, lepton flavours
are distinguishable and  the production of the
lepton asymmetry  should be studied  flavour 
by flavour.  The  baryon asymmetry
calculated in this way is different from 
earlier calculations that considered the production of total lepton number.



\section{Acknowledgements}
It is a pleasure to thank Milla Baldo Ceolin for
inviting me to an interesting, impeccably organised
and enjoyable conference.
These  proceedings are based on work performed in collaboration with
A. Abada,  F.-X. Josse-Michaux, A. Ibarra, M. Losada and
A. Riotto.   Very similar work was performed in parrallel by E. Nardi, Y. Nir,
 J. Racker and E. Roulet. I  thank  AA and ML for
insisting this project was of interest,  AR for
infinite patience and good humour, Nardi {\it et al.}
for  many detailed discussions and a cordial
exchange of drafts, and  as usual Alessandro Strumia (who
did similar work long ago) for the usual irreplaceable
discussions (and disagreements).

\end{document}